%% file: moriond_2002_FDS.tex
\documentclass[11pt]{article}
\usepackage{moriond,epsfig,amssymb}

\input{aliases.tex}

\begin{document}
\vspace*{4cm}
\title{GLUON SATURATION AND \boldmath$S$-MATRIX UNITARITY}

\author{A.~I.~SHOSHI, \underline{F.~D.~STEFFEN}, AND H.~J.~PIRNER}

\address{Institut f\"ur Theoretische Physik, Universit\"at Heidelberg,\\
Philosophenweg 16 {\sl \&}\,19, D-69120 Heidelberg, Germany}

\maketitle\abstracts{The impact parameter dependent gluon distribution
  of the proton $xG(x,Q^2,|\vec{b}_{\!\perp}|)$ is investigated in a
  loop-loop correlation model that respects the $S$-matrix unitarity
  condition in impact parameter space. We find low-$x$ saturation of
  $xG(x,Q^2,|\vec{b}_{\perp}|)$ as a manifestation of $S$-matrix
  unitarity. The integrated gluon distribution $xG(x,Q^2)$ does not
  saturate because of the growth of the effective proton radius with
  decreasing $x$.}

% ___ Introduction _____________________________________________________________
\section{Introduction}
\label{Sec_Introduction}
% ______________________________________________________________________________

The steep rise of the gluon distribution $xG(x,Q^2)$ and structure
function $F_2(x,Q^2)$ of the proton towards small $x=Q^2/s$ is one of
the most exciting results of the HERA
experiments.\cite{Adloff:1997mf+X} As $F_2(x,Q^2)$ is equivalent to
the total $\gamma^* p$ cross section, $\sigma_{\gamma^*
  p}^{tot}(s,Q^2)$, this behavior reflects the rise of
$\sigma_{\gamma^* p}^{tot}(s,Q^2)$ with increasing c.m.\ energy
$\sqrt{s}$ which becomes stronger with increasing photon virtuality
$Q^2$. In hadronic interactions, the rise of the total cross sections
is limited by the Froissart bound, which is a direct consequence of
$S$-matrix unitarity, $SS^{\dagger} = S^{\dagger}S = \Identity$, and
allows at most a logarithmic energy dependence at asymptotic energies.
Analogously, the rise of $\sigma_{\gamma^* p}^{tot}(s,Q^2)$ is
expected to slow down. The microscopic picture behind this slow-down
is the concept of gluon saturation: Since the gluon density in the
proton becomes large at high energies $\sqrt{s}$ (small $x$), gluon
fusion processes are expected to tame the growth of $\sigma_{\gamma^*
  p}^{tot}(s,Q^2)$, and it is a key issue to determine the energy at
which these processes become significant.

In this talk, gluon saturation is considered in an effective loop-loop
correlation model (LLCM) that respects the $S$-matrix unitarity
condition in impact parameter space and allows a unified description
of $pp$, $\gamma^* p$, and $\gamma\gamma$
reactions.\cite{Shoshi:2002in} Concentrating on $\gamma_L^* p$
reactions, the impact parameter dependent gluon distribution of the
proton $xG(x,Q^2,|\vec{b}_{\!\perp}|)$ is computed and found to
saturate in accordance with $S$-matrix unitarity. The presented
results are extracted from Ref.~2 where more details can be found.

% _____________________________________________________________________________
\section{The Loop-Loop Correlation Model}
\label{Sec_The_Model}
% ______________________________________________________________________________

Recently, we have developed a loop-loop correlation model (LLCM) to
compute high-energy hadron-hadron, photon-hadron, and photon-photon
reactions involving real and virtual photons as
well.\cite{Shoshi:2002in} Based on the functional integral approach to
high-energy
scattering,\cite{Nachtmann:1991ua,Dosch:1994ym,Nachtmann:ed.kt} the
$T$-matrix element for elastic $\gamma_L^* p$ reactions with
transverse momentum transfer $\vec{q}_{\perp}$ ($t =
-\vec{q}_{\!\perp}^{\,\,2}$), c.m.\ energy squared $s$, and photon
virtuality $Q^2$ reads
\bea
        \!\!\!\!\!\!\!\!
        T_{\gamma_L^* p}(s,t,Q^2) 
        & \!\!\!\!=\!\!\! &
        2is \!\!\int \!\!d^2b_{\!\perp} 
        e^{i {\vec q}_{\!\perp} {\vec b}_{\!\perp}}
        J_{\gamma_L^* p}(s,|\vec{b}_{\!\perp}|,Q^2)  
\label{Eq_T_gLp_matrix_element} \\        
        \!\!\!\!\!\!\!\!
        J_{\gamma_L^* p}(s,|\vec{b}_{\!\perp}|,Q^2)  
        & \!\!\!\!=\!\!\! & 
        \!\!\int \!\!dz_1 d^2r_1 \!\!\!\int \!\!dz_2 d^2r_2      
        |\psi_{\gamma_L^*}(z_1,\vec{r}_1,Q^2)|^2 |\psi_p(z_2,\vec{r}_2)|^2
        \left(1-S_{DD}({\vec b}_{\!\perp},z_1,{\vec r}_1,z_2,{\vec r}_2)\right)
\label{Eq_model_gp_profile_function}
\eea
where the correlation of two light-like Wegner-Wilson loops, the {\em
  loop-loop correlation function},
\be
        S_{DD}({\vec b}_{\!\perp},z_1,{\vec r}_1,z_2,{\vec r}_2)
        = \Big\langle W[C_1] W[C_2] \Big\rangle_G
        \quad
        \mbox{with}
        \quad
        W[C_{i}] = 
        \inv{3} \Tr\,\Pc
        \exp\!\left[-i g\!\oint_{C_{i}}\!\!\!\!dz^{\mu}
        \G_{\mu}(z) \right]      
\label{Eq_S_DD_def_W[C]_def}
\ee
describes the elastic scattering of two light-like color-dipoles (DD)
with transverse size and orientation ${\vec r}_i$ and longitudinal
quark momentum fraction $z_i$ at impact parameter ${\vec
  b}_{\!\perp}$, i.e., the loops $C_i$ represent the trajectories of
the scattering color-dipoles. For elastic $\gamma_L^* p$ scattering,
the color-dipoles are given by the quark and antiquark in the photon
and in a simplified picture by a quark and diquark in the proton. The
${\vec r}_i$ and $z_i$ distributions of these color-dipoles are given
respectively by the perturbatively derived longitudinal photon wave
function $|\psi_{\gamma_L^*}(z_1,\vec{r}_1,Q^2)|^2$ and the simple
phenomenological Gaussian wave function $|\psi_p(z_2,\vec{r}_2)|^2$.
To account for the non-perturbative region of low $Q^2$ in the photon
wave function, quark masses $m_f(Q^2)$ are used that interpolate
between the current quarks at large $Q^2$ and the constituent quarks
at small $Q^2$.

In contrast to the wave functions, the {\em loop-loop correlation
  function} $S_{DD}$ is universal for $pp$, $\gamma^* p$, and
$\gamma\gamma$ reactions.\cite{Shoshi:2002in} We have computed
$S_{DD}$ in the Berger-Nachtmann approach,\cite{Berger:1999gu} in
which the $S$-matrix unitarity condition is respected as a consequence
of a matrix cumulant expansion and the Gaussian approximation of the
functional integrals. To describe the QCD interaction of the
color-dipoles, we have used the non-perturbative {\em stochastic
  vacuum model}~\cite{Dosch:1987sk+X} supplemented by {\em
  perturbative gluon exchange}. This combination allows us to describe
long and short distance correlations in agreement with lattice
computations of the gluon field strength
correlator~\cite{Bali:1998aj+X} and leads to the static
quark-antiquark potential with color-Coulomb behavior at short
distances and confining linear rise at long
distances.\cite{Euclidean_Model_Applications} Two components are
obtained of which the perturbative ($\pert$) component,
$(\chi^{\pert})^2$, describes two-gluon exchange and the
non-perturbative ($\nprt$) component, $(\chi^{\nprt})^2$, the
corresponding non-perturbative two-point
interaction.\cite{K-Space_Investigations} Ascribing a weak
($\epsilon^{\nprt} = 0.125$) and strong ($\epsilon^{\pert} = 0.73$)
powerlike energy dependence to the non-perturbative and perturbative
component, respectively,
\be
        \left(\chi^{\nprt}(s)\right)^{\!2} \!\!= \left(\chi^{\nprt}\right)^{\!2} \!
        \left(\!\frac{s}{s_0}
        \frac{\vec{r}_1^{\,2}\,\vec{r}_2^{\,2}}{R_0^4}\right)^{\!\!\epsilon^{\nprt}}
        \,\,\mbox{and}\,\,\,\,\,\,\,
        \left(\chi^{\pert}(s)\right)^{\!2} \!\!= \left(\chi^{\pert}\right)^{\!2} \!
        \left(\!\frac{s}{s_0} 
        \frac{\vec{r}_1^{\,2}\,\vec{r}_2^{\,2}}{R_0^4}\right)^{\!\!\epsilon^{\pert}}
        \ ,
\label{Eq_energy_dependence}
\ee
our final result for $S_{DD}$ reads
\be
        S_{DD}
        = \frac{2}{3} 
        \cos\!\left(\frac{1}{3}\chi^{\nprt}(s)\right)
        \cos\!\left(\frac{1}{3}\chi^{\pert}(s)\right)         
        + \frac{1}{3}
        \cos\!\left(\frac{2}{3}\chi^{\nprt}(s)\right)
        \cos\!\left(\frac{2}{3}\chi^{\pert}(s)\right)
        \ .
\label{Eq_S_DD_final_result}
\ee
The cosine functions ensure the unitarity condition in impact
parameter space as they average to zero in the integration over the
dipole orientations at very high energies. In fact, the higher order
terms in the expansion of the cosine functions describe {\em multiple
  gluonic interactions} and are crucial for the saturation of the
impact parameter dependent gluon distribution
$xG(x,Q^2,|\vec{b}_{\!\perp}|)$. Moreover, they tame the growth of the
cross sections at ultra-high energies.\cite{Shoshi:2002in}

% ______________________________________________________________________________
\section{Gluon Saturation}
\label{Sec_Gluon_Saturation}
% ______________________________________________________________________________

Based on the close relationship between the longitudinal structure
function $F_L(x,Q^2)$ and the gluon distribution of the proton
$xG(x,Q^2)$ at small $x$, the {\em impact parameter dependent gluon
  distribution} $xG(x,Q^2,|\vec{b}_{\!\perp}|)$ has been related to
the profile function~\cite{Shoshi:2002in}
$J_{\gamma_L^*p}(s=Q^2/x,|\vec{b}_{\!\perp}|,Q^2)$
\be
        xG(x,Q^2,|\vec{b}_{\!\perp}|) 
        \approx
        1.305\,\frac{Q^2}{\pi^2 \alphaS}\,\frac{\pi}{\alphaEM}\,
        J_{\gamma_L^*p}(0.417 x,|\vec{b}_{\!\perp}|,Q^2)
        \ .
\label{Eq_xg(x,Q^2,b)-J_gLp(x,b,Q^2)_relation}
\ee
Consequently, the shape of $xG(x,Q^2,|\vec{b}_{\!\perp}|)$ is
determined by the profile
function~(\ref{Eq_model_gp_profile_function}), which is a measure for
the blackness or opacity of the interacting particles and, thus, gives
an intuitive geometrical picture for the energy dependence of
$\gamma_L^* p$ reactions: As shown in
Fig.~\ref{Fig_J_gp_(b,s,Q^2)_Fig_xg(x,Q^2,b=0)_vs_x}a, the interacting
particles become blacker and larger with increasing c.m.\ energy
$\sqrt{s}$. At ultra-high energies, $\sqrt{s} \gtsim 10^8\,\GeV$ for
$Q^2 = 1\,\GeV^2$, the opacity saturates at the black disc limit first
for zero impact parameter while the transverse expansion of the
scattered particles continues.
\befig[t] \epsfig{figure=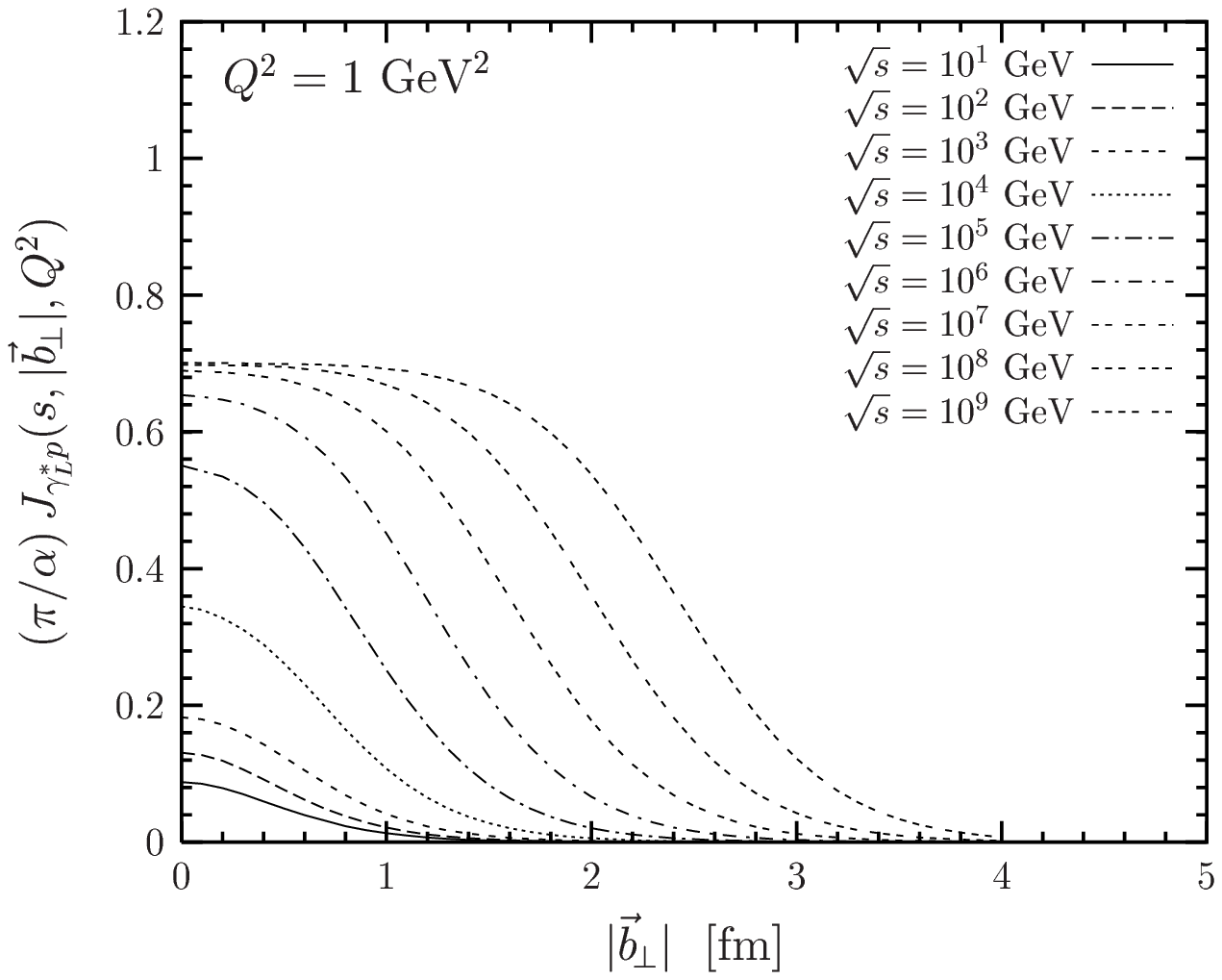,width=7.5cm}\hfill
\epsfig{figure=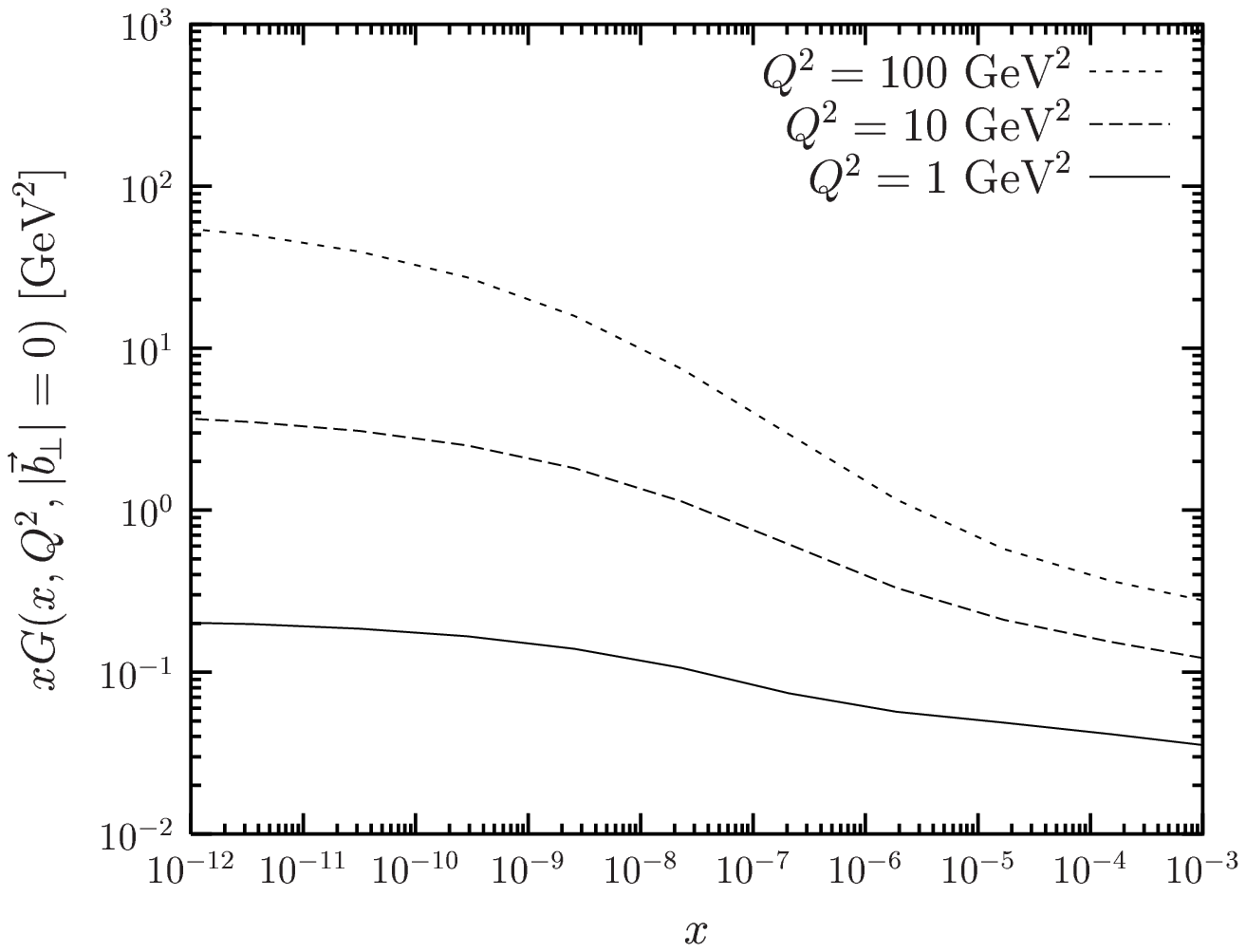,width=7.9cm}
\caption{(a) The profile function
  $({\pi}/\alphaEM)J_{\gamma_L^*p}(s,|\vec{b}_{\!\perp}|,Q^2)$ as a
  function of the impact parameter $|\vec{b}_{\!\perp}|$ at photon
  virtuality $Q^2 = 1\,\GeV^2$ for c.m.\ energies from $\sqrt{s} =
  10\,\GeV$ to $10^9\,\GeV$. (b) The gluon distribution of the proton
  at zero impact parameter, $xG(x,Q^2,|\vec{b}_{\!\perp}|=0)$, as a
  function of $x=Q^2/s$ for $Q^2 = 1,\,10,\,\mbox{and}\,100\,\GeV^2$.}
\label{Fig_J_gp_(b,s,Q^2)_Fig_xg(x,Q^2,b=0)_vs_x}
\end{figure}

In the LLCM with a proton wave function normalized to one, the black
disc limit is given by the normalization of the longitudinal photon
wave function
\be
        J_{\gamma_L^* p}^{max}(Q^2) 
        = \int \!\!dz d^2r |\psi_{\gamma_L^*}(z,\vec{r},Q^2)|^2  
\label{Eq_gp_black_disc_limit}
\ee
and is reached at very high energies when the cosine functions in
$S_{DD}$ average to zero in the $z_i$ and $\vec{r}_i$
integrations~(\ref{Eq_model_gp_profile_function}). Accordingly,
$J_{\gamma^*_L p}^{max}(Q^2)$ induces an upper bound on
$xG(x,Q^2,|\vec{b}_{\!\perp}|)$, the low-$x$ saturation value
\be
        xG(x,Q^2,|\vec{b}_{\!\perp}|)\ \leq \ 
        xG^{max}(Q^2)
        \approx 
        1.305\,\frac{Q^2}{\pi^2 \alphaS}\,\frac{\pi}{\alphaEM}\,
        J_{\gamma^*_L p}^{max}(Q^2)
        \approx 
        \frac{Q^2}{\pi^2 \alphaS}
        \ ,
\label{Eq_low_x_saturation} 
\ee
which is consistent with complementary
investigations~\cite{Mueller:1986wy+X} and indicates strong
color-field strengths $G^a_{\mu \nu} \sim 1/ \sqrt{\alphaS}$ as well.
Since the black disc limit is a rigid unitarity bound for purely
imaginary elastic amplitudes, i.e., obtained directly from the
$S$-matrix unitarity condition in impact parameter space, we conclude
that the low-$x$ saturation of the impact parameter dependent gluon
distribution $xG(x,Q^2,|\vec{b}_{\perp}|)$ is a manifestation of
$S$-matrix unitarity.

In Fig.~\ref{Fig_J_gp_(b,s,Q^2)_Fig_xg(x,Q^2,b=0)_vs_x}b, the
small-$x$ saturation of the gluon distribution at zero impact
parameter $xG(x,Q^2,|\vec{b}_{\!\perp}|=0)$ is illustrated for $Q^2 =
1,\,10,\,\mbox{and}\,100\,\GeV^2$ as obtained in the LLCM. The
saturation occurs at very low values of $x \ltsim 10^{-10}$ for $Q^2
\gtsim 1\,\GeV^2$, where the photon virtuality $Q^2$ determines the
saturation value~(\ref{Eq_low_x_saturation}) and the Bjorken-$x$ at
which it is reached. For larger $Q^2$, the low-$x$ saturation value is
larger and is reached at smaller values of $x$. Moreover, the growth
of $xG(x,Q^2,|\vec{b}_{\!\perp}|=0)$ with decreasing $x$ becomes
stronger with increasing $Q^2$. This results from the stronger energy
increase of the perturbative component in the LLCM, $\epsilon^{\pert}
= 0.73$, that becomes more important with decreasing dipole size. In
conclusion, our approach predicts the onset of the
$xG(x,Q^2,|\vec{b}_{\!\perp}|)$-saturation for $Q^2 \gtsim 1\,\GeV^2$
at $x \ltsim 10^{-10}$ which is far below the $x$-regions accessible
at HERA ($x \gtsim 10^{-6}$) and THERA ($x\gtsim 10^{-7}$).

Note that the $S$-matrix unitarity condition together with
relation~(\ref{Eq_xg(x,Q^2,b)-J_gLp(x,b,Q^2)_relation}) requires the
saturation of the impact parameter dependent gluon distribution
$xG(x,Q^2,|\vec{b}_{\!\perp}|)$ but not the saturation of the
integrated gluon distribution $xG(x,Q^2)$. Indeed, approximating
$xG(x,Q^2,|\vec{b}_{\!\perp}|)$ in the saturation regime by a
step-function,
$xG(x,Q^2,|\vec{b}_{\!\perp}|) 
\approx xG^{max}(Q^2)\,\Theta(\,R(x,Q^2)-|\vec{b}_{\!\perp}|\,)$,
where $R(x,Q^2)$ denotes the full width at half maximum of the profile
function, one obtains
\be
        xG(x,Q^2) 
        \;\approx\;
        1.305\,\frac{Q^2\,R^2(x,Q^2)}{\pi \alphaS}\,
        \frac{\pi}{\alphaEM}\,
        J_{\gamma^*_L p}^{max}(Q^2)
        \;\approx\;
        \frac{Q^2\,R^2(x,Q^2)}{\pi\alphaS}
        \ ,
\label{Eq_xg(x,Q^2)_saturation_regime}
\ee
which does not saturate because of the increase of the effective
proton radius $R(x,Q^2)$ with decreasing $x$. Nevertheless, although
$xG(x,Q^2)$ does not saturate, the saturation of
$xG(x,Q^2,|\vec{b}_{\!\perp}|)$ leads to a slow-down in its growth
towards small $x$.

% ___ Conclusion _______________________________________________________________
\section{Conclusion}
\label{Sec_Conclusion}
% ______________________________________________________________________________

We have computed the impact parameter dependent gluon distribution
$xG(x,Q^2,|\vec{b}_{\!\perp}|)$ from the profile function for
$\gamma^*_L p$ reactions in a loop-loop correlation model (LLCM). The
LLCM combines perturbative and non-perturbative QCD in agreement with
lattice investigations, provides a unified description of $pp$,
$\gamma^* p$, and $\gamma\gamma$ reactions, and respects the
$S$-matrix unitarity condition in impact parameter space. As a
manifestation of $S$-matrix unitarity, we have found low-$x$
saturation of $xG(x,Q^2,|\vec{b}_{\!\perp}|)$ for $Q^2 \gtsim
1\,\GeV^2$ at $x \ltsim 10^{-10}$ but only a slow-down of the
integrated gluon distribution $xG(x,Q^2)$ since the effective proton
radius grows with decreasing~$x$.

% ___ References _______________________________________________________________
\section*{References}
% ______________________________________________________________________________

\end{document}

%% file: aliases.tex
\newcommand{\alphaS}{\alpha_s}
\newcommand{\alphaEM}{\alpha}

\newcommand{\pert}{P}
\newcommand{\nprt}{N\!P}

\newcommand{\be}{\begin{equation}}
\newcommand{\ee}{\end{equation}}
\newcommand{\bea}{\begin{eqnarray}}
\newcommand{\eea}{\end{eqnarray}}
\newcommand{\benn}{\begin{displaymath}}
\newcommand{\eenn}{\end{displaymath}}
\newcommand{\beann}{\begin{eqnarray*}}
\newcommand{\eeann}{\end{eqnarray*}}
\newcommand{\barray}{\begin{array}}
\newcommand{\earray}{\end{array}}
\newcommand{\inv}{\frac{1}}
\newcommand{\gtsim}{\;\lower-0.45ex\hbox{$>$}\kern-0.77em\lower0.55ex\hbox{$\sim$}\;}
\newcommand{\ltsim}{\;\lower-0.45ex\hbox{$<$}\kern-0.77em\lower0.55ex\hbox{$\sim$}\;}

\newcommand{\GeV}{\mbox{GeV}}

\newcommand{\G}{{\cal G}}       % Gluon Field Strength
\newcommand{\Identity}{{1\!\rm l}}% Unit Matrix

\newcommand{\Pc}{{\cal P}}      % Path Ordering
\newcommand{\Tr}{\mbox{Tr}}    % Trace
\newcommand{\befig}{\begin{figure}}
\newcommand{\efig}{\end{figure}}
\newcommand{\betab}{\begin{table}}
\newcommand{\etab}{\end{table}}